\documentclass[11pt,amstex,aps,amsfonts,prsty,preprint]{article} 
\usepackage[dvips]{graphicx}
\usepackage{amssymb}
\usepackage{amsmath}
\usepackage{a4wide}
\usepackage{citesort}
\newcommand{\be}{\begin{equation}}
\newcommand{\ee}{\end{equation}}

\begin{document}

\begin{titlepage}

\begin{center}

{\Large \bf Adsorption  of Reactive Particles on
a Random Catalytic Chain: 
An Exact Solution.}

\vspace{0.1in}

{\Large  G.Oshanin$^{1}$ and S.F.Burlatsky$^{2}$}

\vspace{0.1in}

{$^{1}$ \sl Laboratoire de Physique Th{\'e}orique des Liquides, \\
Universit{\'e} Paris 6, 4 Place Jussieu, 75252 Paris, France}

\vspace{0.1in}

{$^{2}$ \sl United Technologies Research Center,\\ 
United Technologies Corporation,\\
411 Silver Lane, 129-21 East Hartford, CT 06108, USA
}

\vspace{0.1in}

\begin{abstract}
We study equilibrium properties of 
a catalytically-activated  
annihilation $A + A \to 0$  
reaction taking place on a one-dimensional 
chain of length $N$ ($N \to \infty$) 
in which some segments (placed at 
random, with mean concentration $p$) possess 
special, catalytic properties. 
Annihilation 
reaction takes place, as soon as
any two $A$ particles 
land onto two vacant sites at the
extremities of the 
 catalytic segment, or when any 
$A$ particle 
lands onto a vacant site on 
a catalytic segment
while the site at the 
other extremity of this
segment is already occupied 
by another $A$ particle.   
Non-catalytic segments are inert with respect to reaction and 
here two adsorbed $A$ particles harmlessly coexist. 
For both "annealed" and "quenched" disorder in placement of
the catalytic segments, we calculate exactly the disorder-average
pressure per site.
Explicit asymptotic 
formulae for the particle
mean density and the compressibility are also presented. 
\end{abstract}

\vspace{0.1in}
\end{center}

\end{titlepage}

\section{Introduction.}

In many industrial and technological processes the  design of
desired chemicals requires the binding of chemically inactive molecules,
which  recombine only
when some third substance - the catalytic substrate - is present  \cite{1a,7a}.
Within the two past decades much effort has been put  
in understanding of the peculiarities of 
such catalytically-activated reactions (CARs).
On one hand,
much progress was made  
in answering 
the question why and how specific catalytic
 substrates
promote  reactions between chemically 
inactive molecules (see, e.g. Ref.\cite{2a}). 
On the other hand, 
considerable theoretical knowledge was
gained from an extensive study of
 a particular
reaction  -  the CO-oxidation in the presence of 
metal surfaces with  catalytic properties 
\cite{3a} (for a recent review se,, e.g., Ref.\cite{dic}).
While the first
aspect \cite{2a} sheds  light on  catalysation
 mechanisms
 and may allow the
calculation of  $K_{el}$ - 
the rate at which two reactants
react 
being in the vicinity of each other and a specific catalytic substrate,
 the results of Refs.\cite{3a} show
that the mere knowledge 
of $K_{el}$
is not sufficient. As a matter of fact,
Refs.\cite{3a} have substantiated 
the emergence of an 
essentially different behavior as compared to the
predictions of the classical, formal-kinetics  scheme
and have shown that under certain conditions
such
collective phenomena 
as phase transitions
or the formation of
 bifurcation patterns may take place \cite{3a}.
Prior to these works on catalytic systems,  
anomalous behavior was
amply demonstrated in other 
schemes \cite{4a,5a,6a}, involving reactions on contact between two particles 
 at any point of the reaction volume 
(i.e., the "completely" catalytic sysems).
It
was realized \cite{4a,5a,6a} 
that the departure from the text-book, formal-kinetic predictions is
due to 
 many-particle
effects, associated with fluctuations in the spatial
 distribution of the reacting species.  
This suggests that 
similarly to such
 "completely" catalytic
reaction 
schemes,
 the behavior of the 
CARs
 may be influenced
 by many-particle
effects.  

Apart from 
the  many-particle effects, 
behavior of the CARs 
might be affected
by the very 
structure of the catalytic substrate, which is often not
well-defined geometrically, 
but must be viewed as being an assembly
of  mobile or localized
 catalytic
sites or islands, whose spatial distribution 
is complex \cite{1a}.
  Metallic catalysts, for instance, 
are often disordered
compact aggregates, the building blocks 
of which are imperfect crystallites
with broken faces, kinks and steps. 
Usually only the steps are 
active in promoting
the reaction and thus the effective catalytic substrate
is the geometrical pattern formed by these steps.
Another example is furnished by porous materials with convoluted surfaces,
such as, e.g.,
silica, alumina or carbons. Here the effective catalytic
 substrate is also only a portion of the total surface area 
because of the selective participation 
of different surface sites to the reaction 
-  closed pores or pores with very small, bottleneck entrances
are inaccessible 
to many reacting molecules. 
Finally, 
for  liquid-phase 
catalytically-activated reactions the 
catalyst can consist of active  groups attached 
to polymer chains in solution.

Such complex morphologies render the theoretical analysis difficult.
As yet, only empirical approaches have been used
to account for the impact of the geometrical
complexity on  the behavior of the CARs, based mostly 
 on  heuristic concepts of
effective reaction order 
or on phenomenological 
generalizations of the formal-kinetic "law of mass action" 
(see, e.g. Refs.\cite{1a} and \cite{7a} for more details). In this way
the parameters entering the equations 
describing the observables (say, 
the mean particles densities) are fixed by  
fits to experimental data and can deviate from the values
prescribed by
 the stoichiometric relations of the reactions involved. 
The important outcome of such descriptions
is that they provide an evidence  of the existing correlations
in the
morphology of the chemically reactive environment. On the other hand,
their  shortcoming is that they do not explain 
the mechanisms underlying the
anomalous kinetic and stationary behavior. 
In this regard, analytical studies of even somewhat 
idealized or simplified 
models, such as, for instance, 
the ones proposed
in Refs.\cite{3a}, are already 
highly desirable since such studies may
  provide an understanding 
of the  effects of different factors on 
the properties of the CARs.

In this paper we study 
the properties
of catalytically-activated annihilation
$A + A \to 0$ reaction 
in  a simple, one-dimensional
model 
with random distribution 
of the catalyst, appropriate to the 
just mentioned situation 
with the catalytically-activated reactions
 on polymer chains.
More specifically, we consider
here the $A + A \to 0$ reaction 
on a one-dimensional regular lattice which is 
brought in contact with a reservoir of $A$ partilces.
Some portion of the intersite intervals (thick black lines 
in Fig.1) on the regular lattice 
possesses special "catalytic" properties such that
they induce an immediate
 reaction $A + A \to 0$, as soon as
two $A$ particles land onto two vacant sites at the
extremities of the  catalytic segment, or an 
$A$ particle lands onto a vacant site
while the site at the 
other extremity of the
catalytic segment is already occupied 
by another $A$ particle.

\begin{figure}[ht]
\begin{center}
\includegraphics*[scale=0.5]{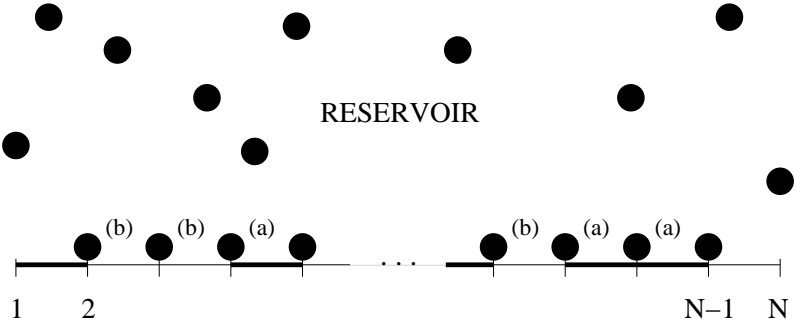}
\caption{\label{Fig1} {\small One-dimensional lattice of 
adsorption sites in contact with a
reservoir. 
Filled circles denote hard-core $A$ particles. 
Thick black lines denote the
segments with catalytic properties. 
(a) denotes a "forbidden" particle
configuration, which corresponds 
to immediate reaction. (b) depicts the situation in which two 
neighboring $A$ particles may harmlessly coexist. 
}}
\end{center}
\end{figure}
  
We present here an exact 
solution of this model 
in two cases - a case when
disorder in placement of the catalytic segments 
can be viewed as $annealed$, and a
more complex situation 
with a
 $quenched$   random distribution of 
the catalytic segments, 
and show that despite the
 apparent oversimplified
nature of the model it exhibits an 
interesting non-trivial behavior. A brief account of these 
results has been presented in our earlier short publication \cite{we}. 
We note finally 
that  
kinetics of $A + A \to 0$ reactions involving diffusive $A$ particles
which react upon encounters on randomly placed catalytic sites 
has been discussed already in Refs.\cite{bur,oshan1} and 
\cite{tox}, and a rather
surprising behavior has been found, especially in low-dimensional systems. 
Additionally, 
steady-state properties of $A + A \to 0$ 
reactions between immobile $A$ particles
with long-range reaction probabilities
in systems with external particles input have been 
presented in Refs.\cite{sander} and \cite{deem}
and revealed non-trivial ordering phenomena with 
anomalous input intensity dependence of the 
mean 
particle density, which agrees with 
early experimental findings \cite{ben}. For completely catalytic 1D 
systems,
kinetics of 
$A + A \to 0$ reactions with immobile $A$ particles 
undergoing cooperative desorption
have been discussed in 
Refs.\cite{nic1,nic2} and \cite{nic3}.  
Exact 
solutions for $A + A \to 0$ reactions in 1D completely catalytic systems
in which
 $A$ particles perform conventional
diffusive or subdiffusive motion have been presented in 
Refs.\cite{lush} and \cite{katja}, respectively.

This paper is structured as follows: in section 2 we define the model and introduce
basic notations. In section 3 we focus on the case of $annealed$ disorder and derive
exact closed-form expressions for the pressure per site, as well as present
explicit asymptotic expansions in powers of the activity for the mean particle density
and the compressibility. 
In section 4 we examine the case of $quenched$ disorder. Here,
we show that the thermodynamic limit result for the disorder-averaged pressure per
site can be obtained very directly by noticing a similarity between the
expressions defining the pressure in the model under study and the Lyapunov exponent
of a product of random two-by-two matrices, obtained by Derrida and Hilhorst
\cite{5}. We also derive an explicit expression obeyed by the averaged
logarithm of the partition function, which is 
valid for any chain's length $N$,
and present its large-$N$ expansion; we show, in particular, 
that the first correction
to the thermodynamic
limit result for the disorder-averaged pressure per site is proportional to the
first negative power of $N$. Explicit asymptotic expansions for the mean particle
density and the compressibility are also derived. Finally, in section 5 we conclude
with a brief summary of our results and discussion.

\section{The model.}

Consider a one-dimensional regular lattice of unit spacing 
comprising $N$ adsorption sites 
in contact with a 
reservoir (vapor phase) of identical, non-interacting hard-core
$A$ particles (see, Fig.1). The reservoir is steadily 
maintained at a constant pressure.

The $A$ particles from
the vapor phase can adsorb onto vacant adsorption sites
and desorb back to the reservoir. 
The occupation of the "i"-th adsorption site is
described by the Boolean variable $n_i$, such that
\begin{equation}
n_i = \left\{\begin{array}{ll}
1,     \mbox{ if the "i"-th site is occupied,} \nonumber\\
0,     \mbox{ otherwise.}
\end{array}
\right.
\end{equation}

Suppose next that some of the segments - intervals between neighboring
adsorption sites possess 
"catalytic" properties (thick black lines 
in Fig.1)
in the
sense that they induce an immediate
 reaction $A + A \to 0$, as soon as
two $A$ particles land onto two vacant sites at the
extremities of the  catalytic segment, or an 
$A$ particle lands onto a vacant site at one extremety of the catalytic segment
while the site at the 
other extremity of this 
segment is already occupied 
by another $A$ particle. Two reacted $A$ particles 
instantaneously leave the lattice (desorb
back to the reservoir). Any two $A$ particle adsorbed at extremities of a 
non-catalytic segment harmlessly coexist.

To specify the positions of the 
catalytic segments, we introduce  the quenched variable 
$\zeta_i$, so that $\zeta_0 = \zeta_N = 0$ and 
\begin{equation}
\zeta_i = \left\{\begin{array}{ll}
1,     \mbox{if the $i$-t interval is catalytic, $i = 1,2, \ldots , N - 1$, } \nonumber\\
0,     \mbox{ otherwise.}
\end{array}
\right.
\end{equation}

Now, for a given distribution of the  catalytic segments, the 
partition function $Z_N(\zeta)$ of 
the system under study can be written as follows: 
\begin{equation}
\label{partition}
Z_N(\zeta) = \sum_{\{n_i\}} 
z^{\sum_{i = 1}^N  n_i} \; \prod_{i = 1}^{N - 1} \Big(1 - \zeta_i \; n_i \; n_{i+1} \Big), 
\end{equation}
where 
the summation $\sum_{\{n_i\}}$ extends over all 
possible configurations
$\{n_i\}$, while 
$z$ denotes the activity,
\begin{equation}
z = \exp(\beta \mu), 
\end{equation}
$\mu$ being the chemical potential, 
which accounts for the reservoir pressure
and for the particles' preference for
adsorption.
Note that $Z_N(\zeta)$ in Eq.(\ref{partition})   
is a functional of the
configuration $\zeta = \{\zeta_i\}$. 

It might be instructive to remark 
that $Z_N(\zeta)$ can be also thought of as a 
one-dimensional version of
models describing adsorption 
of hard-molecules \cite{0,01,001,0001,1,2,3,4},
i.e. adsorption limited by the 
"kinetic" constraint 
that any two of the molecules
can neither occupy the same 
site nor appear on the neighboring sites. The most celebrated 
examples of such models are furnished  by
the so-called "hard-squares" model \cite{0,01,001,0001,1},  
or by the  "hard-hexagons" model first 
solved exactly by Baxter \cite{3}.

These
models ehxibit phase transitions. 
The 
universal classification of phase transitions 
is  
known to depend on the dimensionality, the presence of 
further interactions and the way
in which the lattice can be partitioned into sublattices. 
For bipartite lattices and interactions 
dominated by nearest-neighbor exclusion, the ordering transition is the result
of competition between the two sublattice densities. 
The phase transition is thus associated 
with a breaking of the symmetry between these two sublattices. 
For geometrically more complex 
Baxter's hard-hexagon model, which consists of 
particles with the nearest-neighbor exclusion 
on the triangular lattice, 
the phase transition 
belongs to the three-state Potts 
model universality class, in accordance with the fact 
that the phase transition is 
associated with symmetry breaking involving three competing 
equivalent sublattice densities. 
For more discussion see Ref.\cite{heringa}. 

In our case of the CARs on random catalytic substrates
the nearest-neighbor exclusion 
constraint is
introduced only locally, at 
some specified, randomly distributed intervals. 
Such locally frustrated models of 
random reaction/adsorption
thus represent a natural and meaningful 
generalization
of the well-studied exclusion models over systems with disorder. 
Of course, in this context 
two-dimensional situations are of most interest, but
nonetheless 
it might be instructive to 
find examples of such models
which can be solved exactly in one dimension.

Our main goal here is 
to calculate 
the disorder-average pressure
per site:
\begin{equation}
\label{3}
P_{\infty}^{(quen)} = \frac{1}{\beta} 
\lim_{N \to \infty} \frac{1}{N} \Big<\ln(Z_N(\zeta)) \Big>_{\zeta},
\end{equation}
where the angle brackets with the subscript $\zeta$ here and henceforth 
denote 
averaging over all possible configurations $\{\zeta_i\}$. 
In this case, we suppose that $\zeta_i$ are quenched, independent, randomly distributed
variables with distribution
\begin{equation}
\rho(\zeta) = p \delta(\zeta - 1) + (1 - p) \delta(\zeta).
\end{equation}
As well, we will consider the case when 
the disorder in placement of the catalytic segments can be viewed as
$annealed$ and the mean density of the catalytic segments is equal to $p$; 
in this case, which requires 
much simplier analysis,
the pressure per site is given by
\begin{equation}
\label{8}
P_{\infty}^{(ann)} = \frac{1}{\beta} 
\lim_{N \to \infty} \frac{1}{N}\ln\Big( \Big<Z_N(\zeta) \Big>_{\zeta}\Big),
\end{equation}
We note that such a situation can be realized in practice in case 
when the catalytic agents, modelled here as the catalytic segments, 
diffuse. On the other hand, an assumption of the $annealed$ disorder is 
often used as a meaningful "mean-field" approximation 
for systems with quenched disorder. 
Hence, it might be instructive 
to consider this case in order to check
the behavior provided by such a mean-field approach 
against an exact solution in the quenched disorder case.

Once $P_{\infty}$ are obtained,  
all other pertinent thermodynamic properties can be readily evaluated
by differentiating $P_{\infty}$ with respect to the chemical potential $\mu$; 
in particular,
the disorder-average mean particle density $n$ will be given by
\begin{equation}
\label{dens}
n_{\infty} = \frac{\partial }{\partial \mu} P_{\infty},
\end{equation}
while the compressibility $k_T$
obeys 
\begin{equation}
k_T = \frac{1}{n^2_{\infty}} \frac{\partial n_{\infty}}{\partial \mu}. 
\end{equation}
We set out to show that for both $annealed$ and $quenched$ disorder
cases, 
when $\zeta_i$ are independent, two-state
random variables
all these 
functions can be evaluated explicitly, 
in a closed form. We will distinguish between these two cases 
by assigning  superscripts $(ann)$ and $(quen)$.

To close this section, we display the results 
corresponding to two "regular" cases: namely, when $p = 0$ and $p = 1$, 
which will serve us in what follows as
some benchmarks. In the $p = 0$ all sites are decoupled, and 
one has the trivial Langmuir adsorption results:
\begin{equation}
P^{(lan)}_{\infty} = \frac{1}{\beta} \ln(1 + z), \;\;\; n^{(lan)}_{\infty} = \frac{z}{1 + z}, 
\end{equation}
and
\begin{equation}
\beta^{-1} k_T^{(lan)} = \frac{1}{z}.
\end{equation} 
The "regular" case when $p = 1$ is a bit 
less trivial, but the solution can be still straightforwardly obtained. 
In this case, we have
\begin{equation}
\label{ord}
P^{(reg)}_{\infty} = \frac{1}{\beta} \ln\Big(\frac{\sqrt{1 + 4 z} + 1}{2}\Big), \;\;\; 
n^{(reg)}_{\infty} = 1 - \frac{2 z}{1 + 4 z - \sqrt{1 + 4 z}},
\end{equation}
and
\begin{equation}
\beta^{-1} k_T^{(reg)} = \frac{2 z}{\sqrt{1 + 4 z} (1 + 2 z - \sqrt{1 + 4 z})}.
\end{equation}
Note that in the $p=1$ case (the completely catalytic system) 
the mean particle density tends to $1/2$ as $z \to \infty$ 
(compared to $n_{\infty}^{(lan)} \to 1$ behavior 
observed for the Langmuir case), which means that the 
adsorbent undergoes "ordering" transition
and particles distribution on the lattice becomes 
periodic revealing a spontaneous symmetry breaking between 
two sublattices. 
In the limit $z \to \infty$ 
the compressibility 
vanishes
as $k_T^{(reg)} \propto 1/\sqrt{z}$ compared to the Langmuir behavior 
$k_T^{(lan)} \propto 1/z$.  

\section{Annealed Disorder.}

We start our analysis of the random reaction/adsorption
 model considering first  the situation in which the disorder in placement
of the catalytic segments can be viewed as $\it annealed$. In this case,
the disorder-averaged pressure per site is defined by Eq.(\ref{8}) and thus has
 a more simple form than that in Eq.(\ref{3}),
since we have to perform averaging not 
of the logarithm of the partition function
in Eq.(\ref{partition}) 
but of the partition function itself.

Averaging of the partition function 
in Eq.(\ref{partition})
over the distribution of
the catalytic segments can be performed very directly and
yields
\begin{equation}
\label{partition5}
Z_N = \Big<Z_N(\zeta)\Big>_{\zeta} = \sum_{\{n_i\}} 
z^{\sum_{i = 1}^N  n_i} \; \prod_{i = 1}^{N - 1}
 \Big(1 - p \; n_i \; n_{i+1} \Big).
\end{equation}
Since $(1 - p \; n_i \; n_{i+1}) \equiv 
\exp(\ln(1-p) \; n_i \; n_{i+1})$, 
the result in the latter equation can be thought of
as a partition function of a one-dimensional lattice gas
with nearest-neigbour repulsive interactions of 
amplitude
$\ln(1/(1-p))$. Note that here the original constraint that
no two particle can be located simultaneously
at the extremeties of the catalytic segments 
is replaced by a more tolerant condition 
that the particles may occupy 
neighboring sites anywhere, but the penalty of 
$2 \ln(1-p)$ has to be paid. For any finite $p < 1$ this penalty can be overpassed
by increasing the chemical potential and hence, for large $z$ 
one may thus expect completely different
behavior in the annealed and quenched disorder cases. On the other hand, for $p =1$ this
penalty gets infinitely large and thus $p = 1$ is a special point. 

Now, to find an explicit form 
of 
$Z_N$ we proceed as follows.
Let us first 
introduce an auxiliary, constrained partition 
function of the form
\begin{equation}
Z_N' = \left. Z_N\right|_{ n_{N}  = 1} = z \sum_{\{n_i\}} 
z^{\sum_{i = 1}^{N-1}  n_i} \; \prod_{i = 1}^{N - 2} 
\Big(1 - p \; n_i \; n_{i+1} \Big)  
\Big(1 - p \; n_{N-1} \Big),
\end{equation}
i.e. $Z_N'$ stands for
the partition function of a one-dimensional lattice gas with a nearest-neighbor repulsion 
and  
fixed occupation of the site 
$i = N$, $n_N = 1$. 
Evidently, we have that
\begin{equation}
\label{1}
Z_N = Z_{N-1} + Z_N'. 
\end{equation}
Next, considering two possible values 
of the occupation variable $n_{N - 1}$, 
i.e. $n_{N - 1} = 0$ and $n_{N - 1} = 1$, we find that $Z_N'$ can 
be expressed through $Z_ {N-2}$ and $Z_{N-1}'$
as
\begin{eqnarray}
\label{2}
Z_N' &=&  z \sum_{\{n_i\}} 
z^{\sum_{i = 1}^{N-2}  n_i} \; \prod_{i = 1}^{N - 3} \Big(1 - p \; n_i \; n_{i+1} \Big) 
+ \nonumber\\
&+&  z^2 (1 - p)  \sum_{\{n_i\}} 
z^{\sum_{i = 1}^{N-2}  n_i} \; \prod_{i = 1}^{N - 3} \Big(1 - p \; n_i \; n_{i+1} \Big)  
\Big(1 - p \; n_{N-2} \Big) = \nonumber\\
&=& z Z_{N-2} + z (1 - p) Z_{N-1}'
\end{eqnarray}
Now, recursion in Eq.(\ref{1}) 
allows us to eliminate $Z_N'$ in Eq.(\ref{2}).
 From Eq.(\ref{1}) we have 
$Z_N' = Z_N - Z_{N-1}$, and consequently, we find from
Eq.(\ref{2}) that the unconstrained 
partition function  $Z_N$ in Eq.(\ref{partition5}) obeys the following recursion
\begin{equation}
\label{rec1}
Z_N = \Big(1 + z (1 - p)\Big) Z_{N-1} +
 z p  Z_{N-2},
\end{equation}
which is to be solved subject to evident
 initial conditions
\begin{equation}
\label{bc}
Z_0 \equiv 1 \;\;\; \text{and} \;\;\; Z_1 \equiv 1 + z.
\end{equation}
Solution of the recursion in Eq.(\ref{rec1}) can be readily obtained by
standard means, i.e. evaluating the generating function for $Z_N$, 
$Z_t = \sum_{N = 1}^{\infty} Z_N t^N$, 
 and then inverting it with respect to the variable $t$,
which yields
\begin{equation}
Z_N = \frac{(1 + z p t_+)}{z p t_+ (t_+ - t_-)} t_+^{-N} -  \frac{(1 + z p t_-)}{z p t_- (t_+ - t_-)} t_-^{-N},
\end{equation} 
where
\begin{equation}
t_{\pm} = \pm \frac{1}{2 z p} \sqrt{\Big(1 + z (1 - p)\Big)^2 + 4 z p} - \frac{\Big(1 + z (1 - p)\Big)}{2 z p}
\end{equation}
Noticing next that $t_+ \leq |t_-|$ we find that
 in the annealed disorder case in the thermodynamic limit
the disorder average pressure per site is given by
\begin{equation}
\label{ll}
P_{\infty}^{(ann)} = - \frac{1}{\beta} \ln\Big[ \frac{1}{2 z p} \sqrt{\Big(1 + 
z (1 - p)\Big)^2 + 4 z p} - \frac{\Big(1 + z (1 - p)\Big)}{2 z p}\Big],
\end{equation}
which expression is valid for any $z$ and $p$.

Consider now the asymptotic small-$z$ and large-$z$ behavior of the pressure 
$P_{\infty}^{(ann)}$, the mean density $n_{\infty}^{(ann)}$ and the compressibility
$k_T^{(ann)}$.
Expanding $P_{\infty}^{(ann)}$ in Eq.(\ref{ll})   
into the Taylor series in powers of the activity $z$, we find that
in the small-$z$ limit $P_{\infty}^{(ann)}$ follows:
\begin{equation}
\label{ann}
\beta P_{\infty}^{(ann)} = z - \Big(\frac{1}{2} + p\Big) z + \Big(\frac{1}{3} + 2 p + p^2\Big) z^3 - \Big(\frac{1}{4} + 3 p 
+ \frac{9}{2} p^2 + p^3\Big) z^4 + {\cal O}(z^5).
\end{equation}
Note that $P_{\infty}^{(ann)}$  in Eq.(\ref{ann}) 
reduces to
\begin{equation}
\beta P_{\infty}^{(lan)} =  z - \frac{1}{2} z + \frac{1}{3} z^3 - \frac{1}{4} z^4 + {\cal O}(z^5),
\end{equation}
and 
\begin{equation}
\beta P_{\infty}^{(reg)} = z - \frac{3}{2} z^2 + \frac{10}{3} z^3 - \frac{35}{4} z^4 + {\cal O}(z^5),
\end{equation}
for $p = 0$ and $p = 1$, respectively. 
From Eq.(\ref{ann}) we find that in the annealed disorder case 
in the small-$z$ limit the mean particle density
is given by
\begin{equation}
\label{ann1}
n^{(ann)}_{\infty} = z - \Big(1 + 2 p\Big) z^2 +\Big(1 + 6 p + 3p^2 \Big) z^3 -
\Big(1 + 12 p + 18 p^2 + 4 p^3\Big) z^4 + {\cal O}(z^5),  
\end{equation} 
while the compressibility
obeys
\begin{equation}
\label{ann2}
\beta^{-1} k_T^{(ann)} =  \frac{1}{z} + p \Big(2 - p\Big) z - 4 p z^2 
+ 3 p \Big(2 + 3 p\Big) z^3 - 8 p \Big(1 + 4 p + 2 p^2  \Big) z^4 
+ {\cal O}(z^5).
\end{equation}
We consider next the asymptotic behavior of 
$P_{\infty}^{(ann)}$ in the large-$z$ limit. We notice first that here $p = 1$ is actually
a special point; that is, asymptotic large-$z$ behavior of
 $P_{\infty}^{(ann)}$ is completely different for $p < 1$ and $p = 1$ (completely
catalytic systems). For $p < 1$ and $z \gg (1 - p)^{-2}$, we have that the asymptotic
expansion of $P_{\infty}^{(ann)}$ reads
\begin{eqnarray}
\beta P_{\infty}^{(ann)} &=& \ln\Big((1 - p)^2 z\Big) - \ln(1 - p) + \frac{1}{(1 - p)^2
z} - \frac{(1 + 2 p)}{2 (1 - p)^4 z^2} + \nonumber\\
&+& \frac{(1 + 6 p + 3 p^2)}{ 3 (1 - p)^6 z^3} - \frac{(1 + 12 p + 18 p^2 + 4 p^3)}{4
(1 - p)^8 z^4} + {\cal O}(z^5),  
\end{eqnarray}
while in the regular, completely catalytic case $p = 1$ it follows
\begin{equation}
\beta P_{\infty}^{(ann)} = \beta P_{\infty}^{(reg)} = \frac{1}{2} \ln(z) + \frac{1}{2 z^{1/2}} - \frac{1}{48
z^{3/2}} + \frac{3}{1280 z^{5/2}} + {\cal O}\Big(\frac{1}{z^{7/2}}\Big).
\end{equation}
Consequences of such a difference can be seen in a dramatically different behavior of
the mean particle density. For $p < 1$ and $z \gg (1 - p)^{-2}$ we find
\begin{eqnarray}
\label{123}
n_{\infty}^{(ann)} &=& 1 - \frac{1}{(1 - p)^2 z} + \frac{(1 + 2 p)}{(1 - p)^4 z^2} -
\nonumber\\ 
&-& \frac{(1 + 6 p + 3 p^2)}{ (1 - p)^6 z^3} + \frac{(1 + 12 p + 18 p^2 + 4 p^3)}{
(1 - p)^8 z^4} + {\cal O}(z^5),  
\end{eqnarray} 
while in the regular case $p = 1$ the mean particle density is given by
\begin{eqnarray}
n_{\infty}^{(reg)} = \frac{1}{2} - \frac{1}{4 z^{1/2}} + \frac{1}{32 z^{3/2}} -
\frac{3}{512 z^{5/2}} + {\cal O}(z^{-7/2}).  
\end{eqnarray} 
This signifies, in particular, 
that for $p$ arbitrarily close but not equal to unity, the mean density is equal to
$1$ as $z = \infty$, while for $p$ strictly equal to unity the mean density $n_{\infty}^{(ann)} = 1/2$. 
The behavior of $n_{\infty}^{(ann)}$ as a function of $z$ for 
different values of $p$ is depicted in Fig.3.

In
a similar fashion we find that asymptotic behavior of the compressibility $k_T$ is
very different for $p < 1$ and $p = 1$. For $p < 1$ and $z \gg (1 - p)^{-2}$,
$k_T^{(ann)}$ obeys
\begin{eqnarray}
\beta^{-1} k_{T}^{(ann)} &=& \frac{1}{(1 - p)^2 z} - \frac{4 p}{(1 - p)^4 z^2} +
\nonumber\\ 
&+& \frac{3 p (2 + 3 p)}{ (1 - p)^6 z^3} - \frac{8 p (1 + 4 p + 2 p^2)}{
(1 - p)^8 z^4} + {\cal O}(z^{-5}),  
\end{eqnarray} 
while for $p = 1$ and $z \gg 1$ it follows
\begin{equation}
\beta^{-1} k_{T}^{(reg)} = \frac{1}{2 z^{1/2}} + \frac{1}{2 z} + \frac{3}{16 z^{3/2}} -
 \frac{5}{256 z^{5/2}} + {\cal O}\Big(\frac{1}{z^{7/2}}\Big).
\end{equation}

Finally, we realize that in the $annealed$ disorder case for any fixed $z$ 
the compressibility $k_T^{(ann)}$ appears
to be a $\it non-monotonic$ function of $p$. 
To see this, it suffices to notice that, first,
 $k_T^{(lan)} \leq k_T^{(reg)}$, i.e. for any fixed $z$ the value of 
the compressibility for $p = 0$ is always less or 
equal to its value for $p = 1$.
Second, one readily finds that in the vicinity of $p = 1$ the compressibility $k_T^{(ann)}$ obeys
\begin{equation}
\beta^{-1} k_T^{(ann)} = \beta^{-1} k_T^{(reg)} + \frac{4 z^2}{(1 + 4 z)^{3/2}} (1 - p) + {\cal O}\Big((1 - p)^2\Big),
\end{equation}
i.e. for any $z$ the value $k_T^{(reg)}$ corresponding to $p = 1$ is approached from above.
Consequently, for any fixed $z$ the compressibility $ k_T^{(ann)}$ is a non-monotoneous function of 
the mean density $p$ of the catalytic segments. Behavior of the compressibility $k_T^{(ann)}$ as a function of $p$ 
for several different values of $z$ is presented in Fig.4.

\section{Quenched Disorder.}

We turn now to the more complex situation with a quenched disorder, in which case, in order to 
define the disorder-averaged pressure, 
we have to perform averaging of the logarithm of the partition function in Eq.(\ref{partition}). 
Consequently, here we aim to determine the recursions obeyed by$Z_N(\zeta)$ and 
$\Big< \ln(Z_N(\zeta) \Big>_{\zeta}$.

\subsection{Recursion relations for $Z_N(\zeta)$ and 
$\Big< \ln(Z_N(\zeta) \Big>_{\zeta}$.}

We proceed here along essentially the same lines as in the previous section.
We introduce first a constrained partition 
function of the form
\begin{equation}
Z_N'(\zeta) = \left. Z_N(\zeta)\right|_{ n_{N}  = 1} = z \sum_{\{n_i\}} 
z^{\sum_{i = 1}^{N-1}  n_i} \; \prod_{i = 1}^{N - 2} 
\Big(1 - \zeta_i \; n_i \; n_{i+1} \Big)  
\Big(1 - \zeta_{N - 1} \; n_{N-1} \Big),
\end{equation}
where $Z_N'(\zeta)$ now stands for
 the partition function of a system with  
fixed set $\zeta = \{\zeta_i\}$ and fixed occupation of the site 
$i = N$, $n_N = 1$. 
Similarly to Eq.(\ref{1}), we have that $Z_N(\zeta)$ obeys
\begin{equation}
\label{11}
Z_N(\zeta) = Z_{N-1}(\zeta) + Z_N'(\zeta). 
\end{equation}
Next, considering two possible values 
of the occupation variable $n_{N - 1}$, 
i.e. $n_{N - 1} = 0$ and $n_{N - 1} = 1$, we find that $Z_N'(\zeta)$ can 
be expressed through $Z_ {N-2}(\zeta)$ and $Z_{N-1}'(\zeta)$
as
\begin{equation}
\label{22}
Z_N'(\zeta) =  z Z_{N-2}(\zeta) + z (1 - \zeta_{N-1}) Z_{N-1}'(\zeta),
\end{equation}
which parallels the result in Eq.(\ref{2}).
Eliminating $Z_N'(\zeta)$ in Eq.(\ref{22}), 
we find eventually
that the unconstrained 
partition function  $Z_N(\zeta)$ in Eq.(\ref{partition}) obeys the following recursion:
\begin{equation}
\label{rec11}
Z_N(\zeta) = \Big(1 + z (1 - \zeta_{N - 1})\Big) Z_{N-1}(\zeta) +
 z \zeta_{N - 1}  Z_{N-2}(\zeta),
\end{equation}
which is to be solved subject to the initial conditions in Eq.(\ref{bc}).

A conventional way (see, e.g. Ref.\cite{55,56})
to study linear random three-term recursions is to reduce them 
to random maps by introducing the Ricatti variable of the form 
\begin{equation}
R_{N}(\zeta) = \frac{Z_N(\zeta)}{Z_{N-1}(\zeta)}.
\end{equation}
In terms of this variable Eq.(\ref{rec11}) becomes
\begin{equation}
\label{rec2}
R_N(\zeta) = \Big(1 + z (1 - \zeta_{N - 1})\Big)  +
 \frac{z \zeta_{N - 1}}{R_{N-1}(\zeta)}, \;\;\; \text{with} \;\;\;  R_1(\zeta) \equiv R_1 = 1 + z,
\end{equation}
which represents a random homographic relation.
Once $R_N(\zeta)$ is defined for arbitrary $N$,
the partition function $Z_N(\zeta)$ can be readily determined as the product,
\begin{equation}
Z_N(\zeta) = \prod_{i = 1}^N  R_i(\zeta),
\end{equation}
and hence, the desired disorder-average logarithm of the partition function 
will be obtained as
\begin{equation}
\label{sum}
\Big< \ln Z_N(\zeta)\Big>_{\zeta} = \sum_{i = 1}^N \Big< \ln R_i(\zeta) \Big>_{\zeta}
\end{equation}

Before we proceed further on, some comments on 
the recursion in Eq.(\ref{rec2}) are in order.
We recall first that, 
by definition, each 
quenched random variable $\zeta_{i}$ 
assumes only 
two values - $1$ (with probability $p$) and $0$ 
(with probability $1-p$). 
Hence, we may formally rewrite the random homographic relation in Eq.(\ref{rec2}) as 
\begin{equation}
\label{rec5}
R_i(\zeta) = \left\{\begin{array}{ll}
1 + z/R_{i-1}(\zeta),    \;\;\;  \mbox{$\zeta_{i-1} = 1$, \;\;\; (with probability $p$),} \nonumber\\
1 + z = R_1,  \;\;\;   \mbox{$\zeta_{i-1} = 0$, \;\;\; (with probability $1 - p$).}
\end{array}
\right.
\end{equation}

Note now that recursion schemes of quite a similar form 
have been discussed already in the 
literature in different contexts. 
In particular, two decades ago 
Derrida and Hilhorst \cite{5} (see also Ref.\cite{99} 
for a more general discussion) 
have shown that such recursions occur
in the analysis of 
the Lyapunov exponent $F(\epsilon)$
of the product of random $2 \times 2$ matrices of the form
\begin{equation}
\label{pp}
F(\epsilon) = \lim_{N \to \infty} \frac{1}{N}\Big< \ln\Big({\rm Tr}\Big[\prod_{i=1}^{N} 
\begin{pmatrix}
1 & \epsilon \\
z_i \epsilon & z_i 
\end{pmatrix}
\Big]\Big)\Big>_{\{z_i\}},
\end{equation}
where $z_i$ are independent positive random variables with a given probability distribution $\rho(z)$.  
Equation (\ref{pp}) is related, for instance, to the disorder-average 
free energy of  
an Ising chain with nearest-neighbor interactions in a random
magnetic field, described by the Hamiltonian
\begin{equation}
H' = - J' \sigma_1 \sigma_N - J' \sum_{i = 1}^{N - 1} \sigma_i \sigma_{i+1} - \sum_{i = 1}^N h'_i \sigma_i,
\end{equation}
in which one sets 
$J' = \ln(1/\sqrt{\epsilon})$ and $h'_i = \ln(1/\sqrt{z_i})$.
As noticed  in Ref.\cite{5}, the 
product in Eq.(\ref{pp}) also appears in the solution of a 
two-dimensional Ising model with row-wise random 
vertical interactions \cite{6}, the role of
$\epsilon$ being played by the wavenumber $\theta$. 
The recurence scheme in Eq.(\ref{rec5}) 
emerges also in such 
an interesting context as the problem 
of enumeration of primitive words with 
random errors in the locally free 
and braid groups \cite{100}.   
Some other examples of physical systems 
in which the recursion in Eq.(\ref{rec5}) 
appears 
can be found in \cite{55}. 

Further on, Derrida and Hilhorst \cite{5} 
have demonstrated 
that $F(\epsilon)$ can be expressed as
\begin{equation}
F(\epsilon) = \lim_{N \to \infty} \frac{1}{N} \sum_{i=1}^{N} \Big<\ln R'_i\Big>_{\{z_i\}},
\end{equation}
where  $ R'_i$ are defined through the recursion
\begin{equation}
\label{rec3}
 R'_i = 1 +z_{i-1} + z_{i-1}(\epsilon^2 - 1)/R'_{i-1}, \;\;\; 
\text{with} \;\;\; R'_1 = 1.
\end{equation} 
Moreover,  
they have shown that the model admits an exact solution 
 when 
\begin{equation}
\label{dist}
\rho(z) = (1 - p) \delta(z) + p \delta(z - y),
\end{equation}
i.e. when similarly to the model under study,
 $z_i$ are independent, random two-state 
variables assuming only two values - 
$y$ with probability $p$ and $0$ with probability $1 - p$. 
Supposing that when $i$ increases, a stationary probability
distribution $P(R')$ of the $R'_i$ independent of $i$ exists \cite{8},
Derrida and Hilhorst \cite{5} have found the following exact result:
\begin{eqnarray}
\label{k}
F(\epsilon) &=& p \ln(1 + b) - p (2 - p) \ln(1 + b \frac{y - b}{1 - b y}) + \nonumber\\ 
&+& (1 - p)^2 \sum_{N = 1}^{\infty} p^N \ln(1 + b \Big(\frac{y - b}{1 - b y}\Big)^{N + 1}),
\end{eqnarray}
where
\begin{equation}
\label{k1}
b = 1 + \frac{(1 - y)^2}{2 \epsilon^2 y} 
\Big[1 - \Big(1 + 4 \frac{\epsilon^2 y}{(1 - y)^2}\Big)^{1/2}\Big].
\end{equation}
In particular, Eq.(\ref{k}) shows a striking behavior in 
the $\epsilon \to 0$ limit; in this case,  Derrida and Hilhorst \cite{5} 
have demonstrated 
that for
\begin{equation}
\label{limit}
p y > 1, \;\;\; \text{and} \;\;\; p < 1,
\end{equation}
which implies that $\int \rho(z) \ln(z) \;  < \; 0$, the Lyapunov exponent $F(\epsilon)$ 
defined by Eq.(\ref{k}) exhibits an anomalous, singular behavior
of the form
\begin{equation}
\label{limit1} 
F(\epsilon) \sim \epsilon^{\alpha}, \;\;\; \text{where} \;\;\; \alpha = - \ln(p)/\ln(y).
\end{equation}

We turn now back to our recursion scheme in Eq.(\ref{rec2}) and notice that setting
\begin{equation}
R_i(\zeta) = (1 + z) \; R'_i,
\end{equation}
and choosing
\begin{equation}
\label{def}
y = - \frac{z}{1 + z} = - n^{(lan)}_{\infty}, \;\;\; \text{and} \;\;\; \epsilon^2 = \frac{z}{1 + z} = n^{(lan)}_{\infty},
\end{equation}
makes the recursion schemes in Eqs.(\ref{rec2}) and (\ref{rec3}) identic! 
Consequently, the disorder-average pressure
per site in our random catalytic reaction/adsorption model can be expressed as
\begin{equation}
\label{press}
P_{\infty}^{(quen)} \equiv \frac{1}{\beta} \ln(1 + z) + \frac{1}{\beta} F(\epsilon),
\end{equation}
where $F(\epsilon)$ is the 
Lyapunov exponent of the product of random $2 \times 2$ matrices in Eq.(\ref{pp}), in which
$\epsilon$ and $z_i$ are defined by Eqs.(\ref{dist}) and (\ref{def}). 

Note next that the first 
term on the right-hand-side
of Eq.(\ref{press}) is a trivial Langmuir result for the $p = 0$ case 
(adsorption without reaction) which would entail $n_{\infty}^{(quen)} = z/(1 + z)$.
Hence, all non-trivial, disorder-induced behavior is
 embodied in the Lyapunov exponent $F(\epsilon)$. 
We hasten to remark, however, that 
despite some 
coincidence of results, 
the random reaction/adsorption 
model under study has completely different underlying physics, as compared
to the model studied  by Derrida and Hilhorst \cite{5}. Thus, one would not expect
any singular overall behavior of pressure in
the $\epsilon \to 0$ limit (which corresponds here
to the limit of vanishingly small 
activities $z$ (or $\mu \to - \infty$), and thus pertains  to $n \ll 1$).   
In consequence, here 
$y$ is also dependent on $z$ and $y \to 0$ in the same manner as $\epsilon$. 
Moreover, in our case
$y < 0$, which invalidates the condition in Eq.(\ref{limit}).

\subsection{Disorder-averaged pressure.}

The disorder-averaged pressure per site can be thus readily obtained
from Eqs.(\ref{k}) and (\ref{k1}) by 
defining 
the parameters 
$y$ and $\epsilon$ as prescribed in Eq.(\ref{def}).
This yields the following explicit representation
\begin{eqnarray}
\label{array5}
\beta P_{\infty}^{(quen)} &=& \ln(\phi_z) - (1 - p) \ln\Big(1 - \omega^2\Big) + \nonumber\\
&+& \frac{(1 - p)^2}{p} \sum_{N = 1}^{\infty} p^{N } 
\ln\Big(1 - (-1)^N  \omega^{N + 2}\Big), 
\end{eqnarray}
where
\begin{equation}
\phi_z = \frac{1 + \sqrt{1 + 4 z}}{2},
\end{equation}
and
\begin{equation}
\label{omega}
\omega = \frac{\sqrt{1 + 4 z} - 1}{\sqrt{1 + 4 z} + 1} = z/\phi_z^2 = 1 - \frac{1}{\phi_z}
\end{equation}
Note that $\phi_z$ obeys $\phi_z (\phi_z - 1) = z$ and thus  
for $z = 1$
the $\phi_1$ is just the "golden mean", $\phi_1 = (\sqrt{5} + 1)/2$. Below we will show
why and how this mathematical constant appears here.

On the other hand, the derivation of the 
result in Eq.(\ref{array5}) can be performed in a 
very straightforward manner 
without resorting to the assumption on existence  
of a stationary probability
distribution $P(R')$; 
the 
intermediate steps of such a derivation contain useful formulae, 
which might be helpful
for the understanding 
of the asymptotic behavior of Eq.(\ref{array5}). Since it allows us to answer also
the question of how the thermodynamic limit
is achieved, 
we find it expedient 
to present such a derivation 
here.

We start with calculation of an explicit form of  
$<\ln R_i(\zeta)>_{\zeta}$. To do it, 
it suffices to notice the following two points: First, we notice that
\begin{equation}
R_i(\zeta) = R_i(\zeta_{i-1},\zeta_{i-2},\zeta_{i-3}, \ldots, \zeta_{1}),
\end{equation} 
and 
\begin{equation}
R_{i-1}(\zeta) = R_i(\zeta_{i-2},\zeta_{i-3},\zeta_{i-4}, \ldots, \zeta_{1}),
\end{equation}
i.e. $R_{i-k}(\zeta)$ depends only on $\zeta_{i-k-n}$ with $n = 1,2, \ldots ,i-k-1$
and is independent of $\zeta_{i-k}$.
Second, with probability $1-p$ the Ricatti variable is set equal to 
$1+z$, i.e. to
its initial value $R_1$, which is a non-random function. 
These two observations allow us to work out an explicit 
formula for $<\ln(R_i(\zeta))>_{\zeta}$ which is valid for any $i$.  

Taking the logarithm of both sides of Eq.(\ref{rec2}) and averaging it with respect to the
distribution of  random variables $\zeta_i$, we have:
\begin{equation}
\label{100}
\Big<\ln(R_i(\zeta))\Big>_{\zeta} = 
\Big<\ln\Big(1 +z (1 - \zeta_{i-1}) +z \frac{\zeta_{i-1}}{R_{i-1}}(\zeta) \Big)\Big>_{\zeta}.
\end{equation}
We note next that since $R_{i-1}(\zeta)$ is independent of $\zeta_{i-1}$, 
we can straightforwardly average 
the right-hand-side of Eq.(\ref{100}) with respect to $\zeta_{i-1}$, i.e. 
\begin{eqnarray}
\Big<\ln(R_i(\zeta))\Big>_{\zeta} &=& \Big<\ln\Big(1 +z (1 - \zeta_{i-1}) +
z \frac{\zeta_{i-1}}{R_{i-1}}(\zeta)\Big) \Big>_{\zeta} = \nonumber\\
&=& (1 - p) \ln(1 + z) + p \Big<  \ln\Big(1 + \frac{z}{R_{i-1}(\zeta)}\Big)\Big>_{\zeta} = \nonumber\\
&=& (1 - p) \ln(1 + z) + \nonumber\\
&+& p \Big<  \ln\Big(1 + 
\frac{z}{{\displaystyle 1+z (1 - \zeta_{i-2}) + z \frac{\zeta_{i-2}}{R_{i-2}(\zeta)}}}
 \Big) \Big>_{\zeta}.
\end{eqnarray}
Now, since that $R_{i-2}(\zeta)$ is independent of $\zeta_{i-2}$, we can again 
average over states 
of this variable, which yields
\begin{eqnarray}
\Big<\ln(R_i(\zeta))\Big>_{\zeta} &=& (1 - p) \ln(1 + z) + p \Big<  \ln\Big(1 + 
\frac{z}{{\displaystyle 1+z (1 - \zeta_{i-2}) + z \frac{\zeta_{i-2}}{R_{i-2}(\zeta)}}}
 \Big) \Big>_{\zeta} = \nonumber\\
&=& (1 - p) \ln(1 + z) + p (1 - p) 
 \ln\Big(1 +\frac{z}{{\displaystyle 1 + z}}\Big) +\nonumber\\
&+&  p^2 \Big< 
 \ln\Big(1 + 
\frac{z}{{\displaystyle 1 +  \frac{z}{R_{i-2}(\zeta)}}}
 \Big) \Big>_{\zeta}  = \nonumber\\
&=& (1 - p) \ln(1 + z) + p (1 - p) 
 \ln\Big(1 +\frac{z}{{\displaystyle 1 + z}}\Big) + \nonumber\\
&+&  p^2 \Big< 
 \ln\Big(1 + 
\frac{z}{{\displaystyle 1 +  \frac{z}{\displaystyle 1 +
z (1 - \zeta_{i-3}) +z \frac{\zeta_{i-3}}{R_{i-3}(\zeta)}}}}
 \Big) \Big>_{\zeta}.
\end{eqnarray}
Noticing again that $R_{i-3}(\zeta)$ is independent of $\zeta_{i-3}$ and etc., 
we arrive eventually at the following
explicit representation for $\Big<\ln(R_i(\zeta))\Big>_{\zeta}$:
\begin{equation}
\label{b}
\Big<\ln(R_i(\zeta))\Big>_{\zeta} = (1-p) \sum_{n = 1}^{i-1} p^{n - 1} {\cal F}_n + p^{i-1} {\cal F}_i,
\end{equation}
where the sum on the right-hand-side of Eq.(\ref{b}) is defined for $i \geq 2$ and equals zero otherwise, 
while 
${\cal F}_n$ denote
 natural logarithms of 
the Stieltjes-type continued fractions of the form
\begin{eqnarray}
\label{F}
{\cal F}_1 = \ln(1+z), \;\;\; {\cal F}_2 &=& \ln\Big(\displaystyle 1+\frac{z}{\displaystyle 1 + z}\Big), \;\;\;
{\cal F}_3 = \ln\left(\displaystyle 1+\frac{z}{\displaystyle 1 + \frac{z}{\displaystyle 1 + z}}\right), \nonumber\\
&\cdots& \nonumber\\
{\cal F}_i &=& \ln\left(1 + 
\frac{z}{{\displaystyle 1+\frac{z}{{\displaystyle
1+\frac{z}{{\displaystyle 1+\frac{\cdots}{{\displaystyle
1+z}}}}}}}}
\right).
\end{eqnarray}

To analyze the leading large-$N$ behavior of the disorder-average pressure 
per site
we resort to
 the standard generating function technique \cite{wilf}, often used, in particular,  
in the analysis of peculiar properties of 
different random walks \cite{hughes}. 
Let us define first an auxiliary generating function
\begin{equation}
{\cal R}_t = \sum_{n=1}^{\infty} t^n \Big<\ln(R_n(\zeta)) \Big>_{\zeta}.
\end{equation}
Then, multiplying both sides of Eq.(\ref{b}) by $t^n$ and performing summation, we readily find that
\begin{equation}
{\cal R}_t \equiv \frac{1 - p t}{p (1 - t)} \sum_{n = 1}^{\infty} t^n p^n {\cal F}_n.
\end{equation}
Consequently, the generating function of the averaged logarithm
of the  partition function $Z_N(\zeta)$ obeys:
\begin{eqnarray}
{\cal Z}_t &=& \sum_{N=1}^{\infty} t^N  < \ln Z_N(\zeta)>_{\zeta} = \sum_{N=1}^{\infty} t^N \sum_{n=1}^N \Big<\ln(R_n(\zeta)) \Big>_{\zeta} = \nonumber\\
&=& \frac{1}{1 - t} {\cal R}_t = \frac{1 - p t}{p (1 - t)^2} \sum_{n = 1}^{\infty} t^n p^n {\cal F}_n,
\end{eqnarray}
and hence, the generating function of an average pressure per site,
defined as
\begin{equation}
{\cal P}_t = \frac{1}{\beta} \sum_{N = 1}^{\infty} \frac{t^N}{N}  < \ln Z_N(\zeta)>_{\zeta},
\end{equation}
attains the form
\begin{equation}
{\cal P}_t = \frac{1}{\beta p} \sum_{N = 1}^{\infty} p^N {\cal F}_N \Big(I_N - p I_{N+1}\Big),
\end{equation}
where 
\begin{equation}
I_N = \int^t_0 d\tau \frac{\tau^{N - 1}}{(1 - \tau)^2}.
\end{equation}

Now, in the large-$N$ limit,
the asymptotic 
behavior of the disorder-average pressure $P_N$ per site in a 
finite chain of length $N$ can be obtained very directly from the
 expansion of ${\cal P}_t$ in the vicinity of the closest to the origin 
singular point
\cite{wilf}, i.e.
 $t = 1$. 
Since, in the limit $t \to 1^-$, $I_N$ obeys
\begin{equation}
I_N = \frac{1}{1 - t} + (N - 1) \ln(1 - t) + {\cal O}(1),
\end{equation}
we have that in this limit ${\cal P}_t$ is given by
\begin{equation}
{\cal P}_t = \frac{1}{1 -t} P_{\infty}^{(quen)} + \ln(1 - t) \Big(p \frac{\partial }{\partial p} P_{\infty}^{(quen)}\Big) + {\cal O}(1),
\end{equation}
where 
\begin{equation}
\label{pressure}
P_{\infty}^{(quen)} = \frac{(1 - p)}{\beta p} \sum_{n = 1}^{\infty} p^n {\cal F}_n.
\end{equation}
Consequently,
we find that in the large-$N$ limit $P_N^{(quen)}$ follows
\begin{equation}
\label{111}
P_{N}^{(quen)} = P_{\infty}^{(quen)} - \frac{1}{N}  
\Big(p \frac{\partial }{\partial p} P_{\infty}^{(quen)}\Big) + {\cal O}\Big(\frac{1}{N^2}\Big),
\end{equation}
in which equation $P_{\infty}^{(quen)}$ in Eq.(\ref{pressure})
is the desired thermodynamic limit result for the 
disorder-average pressure per site in the quenched disorder case.
Note that in virtue of the expansion in 
Eq.(\ref{111}), the corrections to the
thermodynamic limit are proportional to 
the first inverse power of the chain length $N$. Note also that since
\begin{equation}
\lim_{n \to \infty} {\cal F}_n = \ln(\phi_z) = \ln\Big(\frac{1 + \sqrt{1 + 4 z}}{2}\Big)
\end{equation}
i.e. ${\cal F}_n$ is the $n$-th approximant 
of $\ln(\phi_z)$,  $P_{\infty}^{(quen)}$ can be thought of as 
the generating function of such approximants. 
One expects then that for $z < 1$  the sequence of approximants 
converges quickly to $\ln(\phi_z)$; 
expanding 
the $n$-th approximant ${\cal F}_n$ into the Taylor series in powers of $z$, one
has that the 
first $n$ terms of such an 
expansion coincide with 
the first $n$ terms
of the expansion of $\ln(\phi_z)$.
Consequently,  ${\cal F}_n$ and ${\cal F}_{n-1}$ differ
 only by terms of order $z^n$, which signifies that convergence is 
good.
 On the other hand,  
for $z \geq 1$ convergence becomes poor 
and one has to seek for a more suitable 
representation. As a matter of fact,  
already for $z = 1$ one has 
that in  
the limit $n \to \infty$ 
the approximant ${\cal F}_n$ tends to  $\ln(\phi_1)$, 
i.e. 
the logarithm of the 
"golden mean", which is known as the 
irrational number worst approximated by rationals. 
Moreover, for $z \to \infty$ 
the convergence is irregular in the sense that
only the approximants with odd numbers show the same large-$z$ behavior as $\ln(\phi_z)$; 
the approximants with even $n$ all tend as $z \to \infty$ to finite values $\ln(n/2 + 1)$ (see, Fig.2).
\begin{figure}[ht]
\begin{center}
\includegraphics*[scale=0.5]{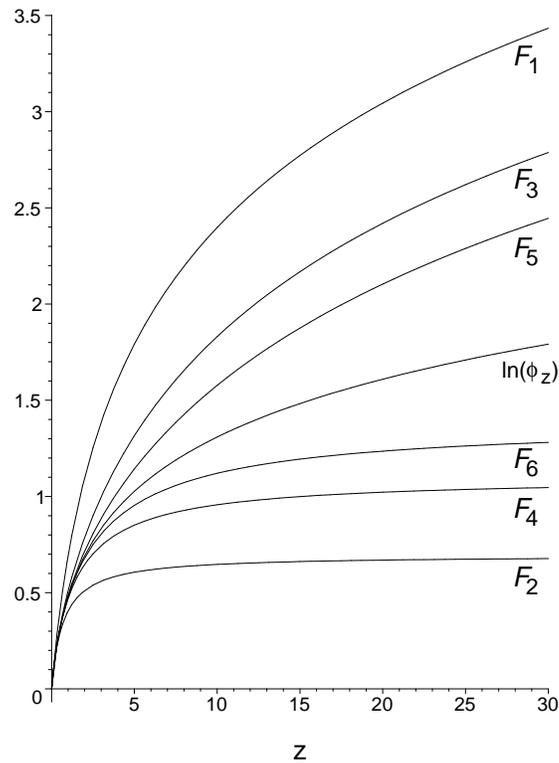}
\caption{\label{Fig2} {\small 
Plot of the approximants ${\cal F}_n$, $n = 1,2,3,4,5$ and $6$, and $\ln(\phi_z)$ 
versus activity $z$. 
}}
\end{center}
\end{figure}
We turn now back to the result in Eq.(\ref{pressure}) aiming to 
find a convenient representation more amenable for further 
analysis.
To do this, let us note that 
${\cal F}_n$ in 
Eq.(\ref{F}) can be expressed as the logarithm of 
the convergents of the Stiltjes-type 
continued fractions:
\begin{equation}
{\cal F}_n = \ln\Big(\frac{K_n(z)}{K_{n-1}(z)}\Big),
\end{equation}
where $K_n(z)$ are polynomials of the activity $z$ 
defined through the 
three-term recursion\footnote{It is straightforward 
to check that the polynomial $K_n(z)$ is just the partition function in Eq.(\ref{partition})
for a chain of length $n$ in the 
completely catalytic $p = 1$ system, i.e. $K_n(z) = Z_n(\zeta \equiv 1)$.}  
\begin{equation}
\label{rec7}
K_n(z) = K_{n - 1}(z) + z K_{n-2}(z), \;\;\; K_0(z) \equiv 1, \;\;\; K_1(z) \equiv 1 + z.
\end{equation}
These polynomials can be, of course, obtained very directly by introducing their 
generating
function, but we can avoid doing it by merely noticing that 
they are simply 
related, in view of the form of the recursion in Eq.(\ref{rec7}), 
to the so-called golden or Fibonacci polynomials $F_{n+2}(x)$ \cite{fibo},
which are 
defined  by the three-term recursion of the form
\begin{equation}
\label{fib}
F_{n+1}(x) = x F_{n}(x) + F_{n-1}(x), \;\;\; F_{1}(x) \equiv 1, \;\;\; F_{2}(x) \equiv x.
\end{equation}
On comparing the recursions in Eqs.(\ref{rec7}) and (\ref{fib}), one infers that
\begin{equation}
K_n(z) = z^{(n+1)/2} F_{n+2}(1/\sqrt{z}).
\end{equation}
Hence, 
the approximant  ${\cal F}_n$ can be expressed as
\begin{equation}
{\cal F}_n =  \frac{1}{2} \ln(z) + \ln\Big(\frac{F_{n+2}(1/\sqrt{z})}{F_{n+1}(1/\sqrt{z})}\Big).
\end{equation}
Note that even at this stage one may understand where from such functions as $\phi_z$
appear
in the expression for the disorder-average pressure in Eq.(\ref{array5}) (first term on the rhs).
The point is that, similarly to the Fibonacci numbers $F_n \equiv F_n(1)$, which 
obey $\lim_{n \to \infty} F_n/F_{n-1} = \phi_1 = (\sqrt{5} + 1)/2$, the ratio of two consequitive golden 
polynomials $F_{n}(1/\sqrt{z})$ and $F_{n-1}(1/\sqrt{z})$
also converges as $n \to \infty$  to a finite limit given by the 
function $\phi_z/\sqrt{z}$. One expects hence that 
the rest of terms on the rhs of Eq.(\ref{array5}) stems from the finite-$n$ effects  and describes 
the relaxation of
the logarithm of  $F_{n}(1/\sqrt{z})/F_{n-1}(1/\sqrt{z})$ to $\ln(\phi_z)$ . 

To determine the relaxation terms, one uses
the standard definition for the Fibonacci polynomials:
\begin{equation}
F_n(x) = \frac{1}{\sqrt{4 + x^2}} \Big[\Big(\frac{x + \sqrt{4 + x^2}}{2}\Big)^n -
 (-1)^n \Big(\frac{2}{x + \sqrt{4 + x^2}}\Big)^n \Big] 
\end{equation}
In virtue of this formula, one finds that the ratio of two consequitive
 golden polynomials obeys
\begin{equation}
\frac{F_{n+2}(1/\sqrt{z})}{F_{n+1}(1/\sqrt{z})} = \frac{\phi_z}{\sqrt{z}} \times 
\left(\frac{\Big[\displaystyle 1 - (-1)^n \omega^{n+2}\Big]}{\Big[\displaystyle 1 
+ (-1)^n \omega^{n+1}\Big]}\right),
\end{equation}
where $\omega$ has been defined in Eq.(\ref{omega}). Consequently, 
we find that the $n$-th approximant 
${\cal F}_n$ is given by
\begin{equation}
\label{last}
{\cal F}_n = \ln(\phi_z) + 
\ln\left(\frac{\Big[\displaystyle 1 - (-1)^n \omega^{n+2}\Big]}{\Big[\displaystyle 1 
+ (-1)^n \omega^{n+1}\Big]}\right),
\end{equation}
where, as we have already remarked, the first term on the rhs of Eq.(\ref{last}) 
corresponds to the limiting form of the approximants, while the second term determines
the relaxation to this limiting form; more specifically, 
to the leading order this relaxation is described 
by an exponential function $\exp(- n \ln(1/\omega))$. Consequently, one expects a fast convergence 
in case when $z$ is small ($\omega$ is small)
and poor convergence when $z \to \infty$ ($\omega \to 1$).
Substituting Eq.(\ref{last}) into Eq.(\ref{pressure}) we recover, 
upon some straightforward algebra,  the result in Eq.(\ref{array5}).

\subsection{Asymptotic behavior of the disorder-average pressure, mean density and the compressibility.}

Consider first the small-$z$
 behavior of the disorder-average pressure per site, 
defined by Eq.(\ref{array5}). As we have already remarked, 
expanding 
the $n$-th approximant ${\cal F}_n$ into the Taylor series in powers of $z$, one
has that the 
first $n$ terms of such an 
expansion coincide with 
the first $n$ terms
of the expansion
\begin{equation}
\ln(\phi_z) =  \ln\Big(\frac{1 + \sqrt{1 + 4 z}}{2}\Big) = - 
\frac{1}{2 \sqrt{\pi}} \sum_{n=1}^{\infty} \frac{(-1)^n \Gamma(n + 1/2)}{\Gamma(n + 1)} \frac{(4 z)^n}{n},
\end{equation}   
which implies that  ${\cal F}_n$ and ${\cal F}_{n-1}$ differ only by terms of order $z^n$ and 
allows to obtain very directly a convergent
small-$z$ expansion of the pressure $P_{\infty}^{(quen)}$. We find then
\begin{eqnarray}
\label{pppp}
\beta P_{\infty}^{(quen)} &=& z - \Big(\frac{1}{2} + p\Big) z^2 + 
\Big(\frac{1}{3} + 2 p + p^2\Big) z^3 - \nonumber\\
&-& \Big(\frac{1}{4} + \frac{7}{2} p + 4 p^2 + p^3 \Big) z^4 + {\cal O}(z^5).
\end{eqnarray}
Consequently, in the small-$z$ limit the mean density obeys
\begin{equation}
\label{nn}
n_{\infty}^{(quen)} = z - (1 + 2 p) z^2 + \Big(1 + 6 p  + 3 p^2\Big) z^3 - 
\Big(1 + 14 p  + 16 p^2 + 4 p^2\Big)z^4  + {\cal O}(z^5),
\end{equation}
while the compressibility $k_T^{(quen)}$ follows
\begin{equation}
\label{kk}
 \beta^{-1} k_T^{(quen)} =  \frac{1}{z} + p (2 - p) z - 4 p (2 - p) z^2 
+  3 p\Big( 8 - p - 2 p^2\Big) z^3 + {\cal O}(z^4).
\end{equation}
Note now that the expressions in Eqs.(\ref{pppp}) to (\ref{kk}) differ from their counterparts
obtained in the $annealed$ disorder case, Eqs.(\ref{ann}), (\ref{ann1}) and (\ref{ann2}), only starting
from the terms proportional to 
the fourth power of the activity $z$. On the other hand, the coefficients 
in the small-$z$ expansion nonetheless coincide with the coefficients in the 
expansions of $P_{\infty}^{(lan)}$ and  $P_{\infty}^{(reg)}$ when we set in Eq.(\ref{pppp}) $p = 0$ or $p = 1$.

Now, we turn to the analysis of the large-$z$ behavior 
which
is a bit more complex than the $z \ll 1$ case 
and requires understanding of the
asymptotic behavior of the sum
\begin{equation}
S = \sum_{N=1}^{\infty} p^N \ln\Big(1 - (-1)^N \omega^{N + 2}\Big)
\end{equation}
entering Eq.(\ref{pressure}). We note first that in this sum 
the behavior of the terms with odd and even $N$ is quite different and we have to consider it separately.

Let
\begin{equation}
S_{odd} = \frac{1}{p} \sum_{N=1}^{\infty} p^{2 N} \ln\Big(1 + \omega^{2 N + 1}\Big)
\end{equation}
denote the contribution of the terms with odd $N$. Note that when $z \to \infty$ (i.e. $\omega \to 1$) 
the sum $S_{odd}$ tends to $p \ln(2)/(1 - p^2)$. The corrections to this limiting behavior can be defined as follows. Expanding
 the logarithm $\ln\Big(1 + \omega^{2 N + 1}\Big)$ into the Taylor series in powers of $\omega$ and 
then, using the definition $\omega = 1 - 1/\phi_z$ and the binomial expansion, we construct a series in the inverse powers of $\phi_z$.
This yields
\begin{eqnarray}
\label{klm}
S_{odd} &=& \frac{p}{1 - p^2} \ln(2) - \frac{1}{2} \frac{p (3 - p^2)}{(1 - p^2)^2} \frac{1}{\phi_z} + \nonumber\\
&+& \frac{1}{8} \frac{p (3 + 6 p^2 - p^4)}{(1 - p^2)^3} \frac{1}{\phi_z^2} + 
 \frac{1}{24} \frac{p (15 + 10 p^2 - p^4)}{(1 - p^2)^3} \frac{1}{\phi_z^3} + {\cal O}\Big(\frac{1}{\phi_z^4}\Big). 
\end{eqnarray}
Note that this expansion is only meaningful when $\phi_z \gg (1 - p)^{-1}$, ($z \gg (1 - p)^{-2}$), which signifies that $p = 1$ is also a special
point for the quenched disorder case.

Further on, plugging into the latter expansion the 
definition of $\phi_z$, $\phi_z = (1 + \sqrt{1 + 4 z})/2$, 
we obtain the following expansion in the inverse powers of the activity $z$:
\begin{eqnarray}
\label{mmm}
S_{odd} &=& \frac{p}{1 - p^2} \ln(2) - \frac{p}{2} \frac{(3 - p^2)}{(1 - p^2)^2} \frac{1}{z^{1/2}} + \nonumber\\
&+& \frac{p}{8} \frac{(9 - 2 p^2 + p^4)}{(1 - p^2)^3} \frac{1}{z} + 
 \frac{p}{48} \frac{(3 - 4 p^2 + p^4)}{(1 - p^2)^3} \frac{1}{z^{3/2}} + {\cal O}\Big(\frac{1}{z^2}\Big). 
\end{eqnarray}

\begin{figure}[ht]
\begin{center}
\includegraphics*[scale=0.5]{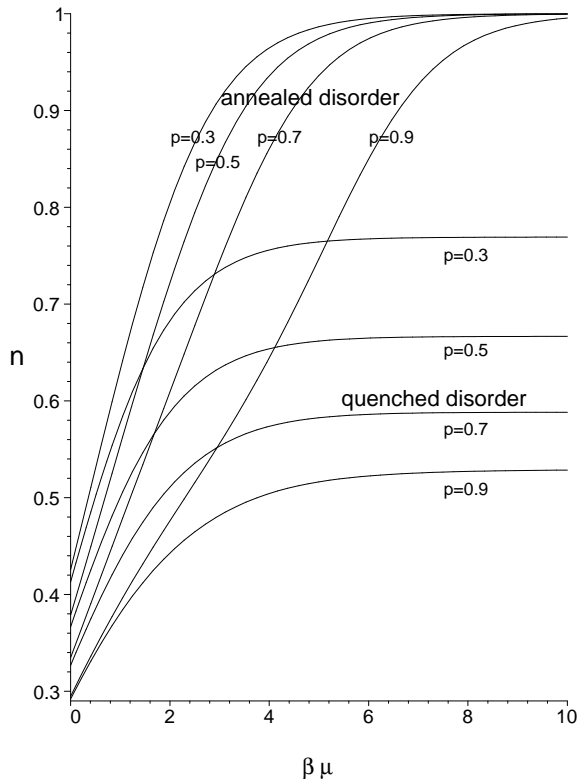}
\caption{\label{Fig3} {\small The mean density of adsorbed particles versus the chemical 
potential $\beta \mu$
for the annealed (curves tending to unity) and quenched disorder case for different values 
of the mean density $p$ of the catalytic segments.
}}
\end{center}
\end{figure}

\begin{figure}[ht]
\begin{center}
\includegraphics*[scale=0.5]{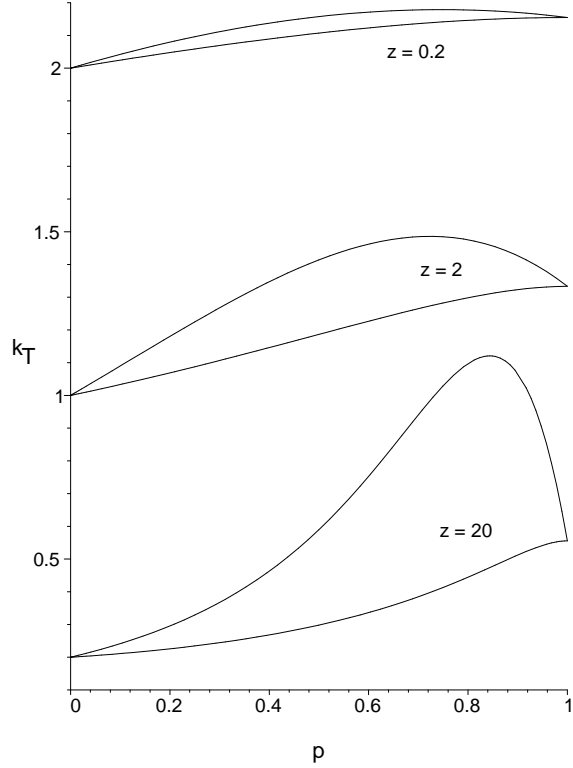}
\caption{\label{Fig4} {\small The compressibility $\beta^{-1} k_T$ versus the mean density $p$ 
of the catalytic segments for several values of the activity $z$, $z = 0.2, 2$ and $z = 20$. Upper, 
non-monotoneous curves show the behavior of $\beta^{-1} k_T$ in the $annealed$ disorder case, while
the lower curves correspond to the solution in the $quenched$ disorder case.
}}
\end{center}
\end{figure}

Consider next the sum
\begin{equation}
\label{even}
S_{even} =  \sum_{N=1}^{\infty} p^{2 N} \ln\Big(1 - \omega^{2 N + 2}\Big),
\end{equation}
which represents the contribution of terms with even $N$. Note 
that in contrast to the behavior of $S_{odd}$, the sum in Eq.(\ref{even})
diverges when $z \to \infty$ ($\omega \to 1$). Since $1 - \omega^{2 N + 2} \sim 1 - \omega$ for $\omega \to 1$, we have
that in this limit
\begin{equation}
\label{even2}
S_{even} \sim \frac{p^2}{1 - p^2} \ln(1 - \omega)
\end{equation} 
To obtain several correction terms we make use of one of Gessel's expansions
\cite{gessel}:
\begin{equation}
\label{gessel}
\ln\Big(\frac{2 (N + 1) x}{1 - (1 - x)^{2 N + 2}}\Big) = \sum_{k = 1}^{\infty} g_k(2 N + 2) \frac{(-1)^k x^k}{k},
\end{equation}
where $g_k(2 N + 2)$ are the Dedekind-type sums of the form
\begin{equation}
g_k(2 N + 2) = \sum_{\zeta^{2 N + 2} = 1, \zeta \neq 1} \frac{1}{\Big(\zeta - 1\Big)^{k}},
\end{equation}  
where the summation extends over all $\zeta$ being the $(2 N + 2)$-th roots of unity (with $\zeta = 1$ excluded).
As shown in Ref.\cite{gessel}, the weights $g_k(2 N + 2)$ are polynomials in $N$ 
of degree at most $k$ with rational coefficients; first few values
of  $g_k(2 N + 2)$ are:
\begin{eqnarray}
\label{g-s}
g_1(2 N + 2) &=& - (2 N + 1)/2, \nonumber\\
g_2(2 N + 2) &=& - (2 N + 1) (2 N - 3)/12, \nonumber\\
g_3(2 N + 2) &=&   (2 N + 1) (2 N - 1)/8, \nonumber\\
g_4(2 N + 2) &=&   (2 N + 1) (8 N^3 + 28 N^2  - 186 N + 45)/720.
\end{eqnarray}  
Now, setting $x = 1/\phi_z$ in the expansion in Eq.(\ref{gessel}), plugging it to Eq.(\ref{even}) 
and performing summations over $N$, 
we find that $S_{even}$ can be written down as
\begin{equation}
S_{even} = - \frac{p^2}{1 - p^2} \ln(\phi_z) + \frac{p^2}{1 - p^2} \ln(2) + s_p - \sum_{k=1}^{\infty} G_k(p) \frac{(-1)^k}{k \phi_z^k},
\end{equation}
where $s_p$ is an infinite series of the form \footnote{Note that $s_p$ shows a non-analytic behavior when $p \to 1$. This function can be represented as
\begin{equation}
s_p = - \frac{1}{1 - p^2} \ln(1 - p^2) - \frac{p^2}{1 - p^2} \sum_{n = 2}^{\infty} \frac{(-1)^n}{n} \Phi(p^2,n,1),
\end{equation}
where $\Phi(p^2,n,1)$ are the Lerch transcedents, $\Phi(p^2,n,1) = \sum_{l = 0}^{\infty} (1 + l)^{-n} p^{2 l}$. It is 
straightforward to find then that $s_p =  - \frac{1}{1 - p^2} \ln(1 - p^2) - \frac{\gamma}{1 - p^2} + {\cal O}(\ln(p))$,
where $\gamma$ is the Euler constant.}
\begin{equation}
s_p = \sum_{N = 1}^{\infty} p^{2 N} \ln(N + 1), 
\end{equation}
while $G_k(p)$ are the generating functions of the polynomials $g_k(2 N + 2)$:
\begin{equation}
G_k(p) = \sum_{N=1}^{\infty} g_k(2 N + 2) p^{2 N} 
\end{equation}
Inserting next the definition of $\phi_z$, we find the following explicit asymptotic expansion
\begin{eqnarray}
\label{j2}
S_{even} &=& - \frac{1}{2} \frac{p^2}{1 - p^2}  \ln(z) + \frac{p^2}{1 - p^2} \ln(2) + s_p + \nonumber\\
&-& \frac{p^2 (2 - p^2)}{(1 - p^2)^2} \frac{1}{z^{1/2}} + \frac{p^2 (21 - 18 p^2 + 5 p^4)}{24 (1 - p^2)^3} \frac{1}{z} +
\frac{p^2 (2 - p^2)}{24 (1 - p^2)^2} \frac{1}{z^{3/2}} + {\cal O}\Big(\frac{1}{z^{2}}\Big).
\end{eqnarray}
Finally, 
combining the expansions in Eqs.(\ref{pressure}), (\ref{mmm}) and (\ref{j2}), we find the desired large-$z$ expansion 
for the disorder-averaged pressure $P_{\infty}^{(quen)}$:
\begin{eqnarray}
\label{presq}
\beta P_{\infty}^{(quen)} &=& \frac{1}{1 + p} \ln(z) - \frac{(1 - p)^2}{(1 + p)} \ln(2) + \nonumber\\ 
&+& \frac{(1 - p)^2}{p} s_p + \frac{1}{6} \frac{6 + 3 p - p^3}{(1 + p)^2 (1 - p^2)} \frac{1}{z} +
 {\cal O}\Big(\frac{1}{z^2}\Big).
\end{eqnarray}
Note that $P_{\infty}^{(quen)}$ in Eq.(\ref{presq}) shows a completely different behavior 
compared to its counterpart in the annealed disorder case already in the leading term
in the large-$z$ expansion. Note also that here $p = 1$ appears to be a special point and thus the expansion in 
Eq.(\ref{presq}) becomes meaningless for $p = 1$. As a matter of fact, for $p$ arbitrarily close but less than unity
one has intervals which are devoid of the catalytic segments. Contribution of such intervals to the overall  
disorder-average pressure is of a Langmuir-type and vanishes only when $p$ is strictly equal to unity, 
which implies that also here $p = 1$ is a special point.

We find next that for $z \gg (1 - p)^{-2}$ the mean particle density obeys
\begin{equation}
\label{densii}
n_{\infty}^{(quen)} = \frac{1}{1 + p} - \frac{1}{6} \frac{6 + 3 p - p^3}{(1 + p)^2 (1 - p^2)} \frac{1}{z} + {\cal O}\Big(\frac{1}{z^2}\Big),
\end{equation}
i.e. contrary to the behavior of the mean particle density in the annealed disorder case, Eq.(\ref{123}),
$n_{\infty}^{(quen)}$ tends towards  a constant value $1/(1 + p)$, which depends 
on $p$ and coincides with the corresponding
values $n^{(lan)} = 1$ and $n^{(reg)} = 1/2$ for $p = 0$ and $p = 1$. Behavior of the mean density versus the chemical potential $\mu$
for the annealed and quenched disorder cases is presented in Fig.3. 

Finally, from Eq.(\ref{densii}) we find that the compressibility $k_T^{(quen)}$ admits the following form:
\begin{equation}
\beta^{-1} k_T^{(quen)} = 
\frac{1}{6} \frac{6 + 3 p - p^3}{(1 + p) (1 - p^2)} \frac{1}{z} +
\frac{1}{36} \frac{p \Big(6 + 3 p - p^3\Big)^2}{(1 + p)^2 (1 - p^2)^2} \frac{1}{z^2} 
+ {\cal O}\Big(\frac{1}{z^3}\Big),
\end{equation}
which expansion also holds in the asymptotic limit $z \gg (1 - p)^{-2}$.

\section{Conclusions.}

To conclude, in this paper we have presented an exact solution of a random reaction/adsorption model, 
appropriate to the 
 situations 
with the catalytically-activated reactions
on polymer chains containing randomly placed catalytist.
More specifically, we have considered
here the $A + A \to 0$ reaction 
on a one-dimensional regular lattice which is brought in contact with a reservoir of $A$ partilces.
Some portion of the intersite intervals  on the regular lattice 
was supposed to possess special "catalytic" properties such that
they induce an immediate
 reaction $A + A \to 0$, as soon as
two $A$ particles land onto two vacant sites at the
extremities of the  catalytic segment, or an 
$A$ particle lands onto a vacant site
while the site at the 
other extremity of the
catalytic segment is already occupied 
by another $A$ particle. 
For two different cases; namely, when
disorder in placement of the catalytic segments can be viewed as $annealed$, and a
more complex situation 
with a
 $quenched$   random distribution of 
the catalytic segments, 
we have determined exactly the disorder-averaged pressure per site. 
For the annealed disorder case 
such a pressure 
has been found in a closed form 
and explicit asymptotic expansions in powers of the activity for the mean particle density
and the compressibility have been obtained. In the case of $quenched$ disorder
we have shown that the thermodynamic limit result for the disorder-averaged pressure per
site can be obtained very directly by noticing a similarity between the
expressions defining the pressure in the model under study and the Lyapunov exponent
of a product of random two-by-two matrices, obtained by Derrida and Hilhorst
\cite{5}. We have also derived an explicit expression obeyed by the averaged
logarithm of the partition function, which is 
valid for any chain's length $N$. From this expression we have constructed
the large-$N$ expansion and have
shown, in particular, 
that the first correction
to the thermodynamic
limit result for the disorder-averaged pressure per site is proportional to the
first negative power of $N$. The leading term in this expansion coincides
with the one found from the analysis by Derrida and Hilhorst. 
Explicit asymptotic expansions for the mean particle
density and the compressibility were also derived. 
We have demonstrated that for low activities in the annealed and quenched disorder cases 
the coefficients in the
corresponding expansions of the pertinent
parameters in the Taylor series in powers of $z$ 
coincide up to the order $z^3$ and slightly deviate from each other starting from the fourth order. 
On the other hand, expansions in inverse powers of $z$ (large-$z$ behavior)
are different already in the leading order. Most spectacular difference
between the annealed and quenched disorder case have been observed in the 
behavior of the compressibility:
in the annealed disorder case it appears to be a non-monotoneous
function of the mean density $p$ of the catalytic segments, while in the quenched
disorder case it is a monotoneously increasing function of $p$.

\end{document}